\documentclass[11pt]{article}
\newcommand{\BE}{\begin{equation}}
\newcommand{\EE}{\end{equation}}
\newcommand{\BA}{\begin{eqnarray}}
\newcommand{\EA}{\end{eqnarray}}

\begin{document}
\begin{titlepage}

\vspace*{1mm}
\begin{center}

            {\LARGE{\bf Long-wavelength excitations of 
Higgs condensates } }

\vspace*{14mm}
{\Large  M. Consoli }
\vspace*{4mm}\\
{\large
Istituto Nazionale di Fisica Nucleare, Sezione di Catania \\
Corso Italia 57, 95129 Catania, Italy}
\end{center}
\begin{center}
{\bf Abstract}
\end{center}

Quite independently of the Goldstone phenomenon, 
recent lattice data suggest the existence of gap-less modes in the
spontaneously broken phase of a $\lambda \Phi^4$ theory.
This result is a direct consequence of the quantum nature of the
`Higgs condensate' that cannot be treated as a purely
classical c-number field.
\end{titlepage}

Spontaneously broken $\lambda \Phi^4$ theories
have been used to fix the ground state in many
quantum field theoretical models, including the Standard Model of electroweak
interactions. Traditionally, 
with the notable exception of the Coleman-Weinberg \cite{CW} paper, 
the symmetry breaking mechanism has been searched into 
a double-well classical potential with perturbative loop
corrections. At the same time, in the absence of the Goldstone phenomenon, 
i.e. for a one-component theory, the particle content of the broken phase
is represented as a single massive field, the
Higgs boson. In this 
picture, the lowest excitations of the theory should have an energy spectrum
of the form
$\tilde{E}({\bf{p}})= \sqrt{ {\bf{p}}^2 + M^2_h}$ so that
the `Higgs mass' $M_h$ 
coincides with $\tilde{E}(0)$, the energy-gap of the broken phase. 

A possible objection to the above simple description can arise after having
tried to represent spontaneous symmetry breaking as a real
`condensation' phenomenon \cite{mech}. Although the nature of the elementary 
condensed quanta is not precisely known, 
it is conceivable that, beyond the simplest approximation where
the `Higgs condensate' is treated as a classical c-number field, there may be
collective effects that can change qualitatively the energy spectrum in the 
long-wavelength region. As an example, let us consider
Landau's quasi-particle picture. This
 has been proved to be very useful to describe the elementary excitations
for a large variety of many-particle
systems. However, there are cases where the leading term
in the energy spectrum is {\it not} proportional to $ {\bf{p}}^2$ when 
${\bf{p}} \to 0$. In these situations, defining an effective
quasiparticle mass as for instance (`NR'= Non-Relativistic)
\BE
\label{meff}
   {{1}\over{2m_{\rm eff} }}= \lim_ 
{ {\bf{p}} \to 0 } 
{{ \partial E_{\rm NR}({\bf{p}})  }\over{ \partial {\bf{p}}^2  }}
\EE
is misleading. For instance, by using Eq.(\ref{meff}) in a system 
that has a phonon excitation branch $\sim |{\bf {p}}|$ 
for ${ {\bf{p}} \to 0 } $ we would find an `effective mass' that 
becomes smaller and smaller in the limit of a vanishing 3-momentum. 
The same may occur in
the broken phase of $\lambda\Phi^4$ theories where the validity of the
identification 
$M_h=\tilde{E}(0)$ depends on the form of the energy spectrum for
${\bf{p}} \to 0$.

The possibility that the Higgs mass
$M_h$ differs non-trivially from the energy-gap of the broken phase
has been objectively addressed \cite{cea2} with lattice simulations.
In this case one can precisely measure the exponential decay
of the connected correlator at various values of $|{\bf{p}}|$ and
determine the energy spectrum of the cutoff theory
$\tilde{E}({\bf{p}})$ by comparing with (the lattice version of)
$\sqrt{ {\bf{p}}^2 + M^2_h}$.

To exclude possible uninteresting effects, 
one should preliminarly perform the same analysis in the symmetric phase and 
compare the corresponding measured energy spectrum 
$E({\bf{p}})$ with the form
$\sqrt{ {\bf{p}}^2 + m^2}$. In the symmetric phase
the lattice data \cite{cea2}
give $E(0)=m$ to very high accuracy as expected.
However, in the broken phase, 
$\tilde{E}^2({\bf{p}})- {\bf{p}}^2$ turns out to depend on 
$|{\bf{p}}|$ when
${\bf{p}} \to 0$ and therefore the attempt to extract $M_h$ from the low-momentum
data becomes problematic. If, on the other hand, $M_h$ is consistently
extracted from those higher-momentum data where 
$\tilde{E}^2({\bf{p}})- {\bf{p}}^2$ does {\it not} depend on 
$|{\bf{p}}|$, then one finds
$\tilde{E}(0)< M_h$ with a discrepancy between the two values that 
seems to increase when taking the continuum limit. 

Moreover, new data \cite{cea3} show that, for the same lattice
action, by increasing the lattice size one finds 
smaller and smaller values of the energy-gap
$\tilde{E}(0)$.
Namely, by using the same lattice parameters that 
on a $20^4$ lattice give \cite{jansen}
$\tilde{E}(0)=0.3912 \pm 0.0012$, one finds \cite{cea3}
$\tilde{E}(0)=0.3791 \pm 0.0035$
on a $24^4$ lattice, 
$\tilde{E}(0)=0.344\pm 0.008$
on a $32^4$ lattice and
$\tilde{E}(0)=0.298\pm 0.015$
on a $40^4$ lattice. Therefore, differently from $M_h$ which, being extracted from
the higher-momentum part of the energy spectrum, shows no dependence on the
lattice volume (see Table 2 and Fig.8 of \cite{cea2}), 
the energy-gap in the broken phase is an infrared-sensitive quantity
that become smaller and smaller by increasing the lattice size and
may even vanish in the infinite-volume limit. 

An indication in this sense has been obtained in ref.\cite{tadpoles} 
by studying the $p \to 0$ limit
of the propagator $G(p)$ 
after re-summing the infinite series of the 
zero-momentum tadpole graphs in a given background field.
 Indeed, the tadpole subgraphs are attached to
the other parts of the diagrams through zero-momentum propagators
and can be considered a manifestation of the quantum nature of the 
scalar condensate. In this case, besides the
usual massive solution $G^{-1}(0)=M^2_h$, one finds gap-less solutions
$G^{-1}(0)=0$ that would not exist otherwise. 

In this Letter, we shall provide another argument in favour of the existence
of such gap-less modes. Our discussion is based on the results obtained by
Ritschel \cite{ritschel} for the effective potential
that we shall first briefly review. 

There are two basically different definitions of the effective potential. 
A first definition is $V_{\rm eff}(\phi)\equiv V_{\rm LT}(\phi)$, i.e.
the Legendre transform (`LT') of the generating functional for connected 
Green's function. In this way $V_{\rm LT}(\phi)$ is rigorously
convex downward. For this reason, it is not the same
thing as the usual non-convex (`NC') effective potential
$V_{\rm eff}(\phi)\equiv V_{\rm NC}(\phi)$ that is found in the ordinary
loop expansion. Moreover, in the presence of spontaneous symmetry breaking, 
$V_{\rm LT}(\phi)$ is {\it not} 
an infinitely differentiable function of $\phi$ \cite{syma}, differently from
$V_{\rm NC}(\phi)$. As shown in ref.\cite{ritschel}, 
the difference between $V_{\rm LT}(\phi)$ and
$V_{\rm NC}(\phi)$ amounts to include the
quantum effects of the zero-momentum mode that cannot be
treated as a purely classical background but requires
one more functional integration in field space.

In the following, we shall assume that
$V_{\rm NC}(\phi)$ has just a pair $\pm v$ of 
degenerate absolute minima (the generalization to the more complex situation
of several competing minima can be done by using Ritschel's formulas \cite
{ritschel}). In addition, we shall assume that $\phi=\pm v$ define the
correct normalization of the background field 
such that the Higgs mass is defined by the quadratic shape of 
$V_{\rm NC}(\phi)$ at the minima, namely
\BE
\label{newident}
       \left. \frac{ d^2 V_{\rm NC}}{d \phi^2} 
\right|_  {  \phi=\pm v   } \equiv M^2_h
\EE
We shall now use 
 Ritschel's results \cite{ritschel} for a space-time constant source $J$ 
where, after integrating 
out all non-zero quantum modes, 
the generating functional is given by
\BE
\label{zetaj}
      Z(J)= \int^{+ \infty}_{-\infty} d\phi~ \exp [-\Omega
(V_{\rm NC}(\phi) - J\phi)]
\EE 
In Eq.(\ref{zetaj}) 
$\Omega$ denotes the four-dimensional space-time volume and
we can define an associated density $w(J)$ as 
\BE
\label{wj}
\ln {{Z(J)}\over{Z(0)}} \equiv \Omega w(J)
\EE
In the saddle-point approximation, valid for $\Omega \to \infty$, we get
\BE
\label{saddle}
w(J)= {{J^2 }\over{2 M^2_h}} + 
{{\ln \cosh(\Omega J v)}\over{\Omega}}
\EE
and
\BE
\label{phij2}
\varphi={{dw}\over{dJ}}= {{J}\over{M^2_h}} + v \tanh(\Omega J v)
\EE
\BE
\label{GJ}
G_J(0)={{d\varphi}\over{dJ}}=
{{1}\over{M^2_h}} + {{\Omega v^2}\over{\cosh^2(\Omega Jv)}}
\EE
Notice that only retaining the full $J-$dependence in Eq.(\ref{saddle}) 
one can fulfill the basic symmetry property
\BE
     \varphi(-J)=-\varphi(J)
\EE
To determine the zero-momentum propagator in a given background $\varphi$, 
we should now invert $J$ as a function of $\varphi$ from Eq.(\ref{phij2})
and replace it in (\ref{GJ}). However, 
being interested in the limit $J \to 0$ it is easier to look for the
possible limiting behaviours of (\ref{GJ}). 

Since both $J$ and $\Omega$ are dimensionful quantities, it is convenient 
to introduce dimensionless variables
\BE
\label{exj}
        x\equiv \Omega J v
\EE    
and
\BE
\label{uai}
          y\equiv \Omega v^2 M^2_h
\EE
so that Eqs.(\ref{phij2}) and (\ref{GJ}) become
\BE
\label{phij3}
   \varphi = v~ [ {{x}\over{y}} + \tanh(x)]
\EE
and
\BE
\label{GJ2}
G_J(0) = {{1}\over{M^2_h}}~[1 + {{y}\over{\cosh^2(x)}}]
\EE
Therefore, the two limits 
$\Omega \to \infty$ and $J \to \pm 0$ correspond to various paths in the 
two-dimensional space $(x,y)$. The former gives simply $y \to \infty$.
The latter, on the other hand, is equivalent to $ {{x}\over{y}} \to \pm 0$ since
\BE
{{J }\over{M^2_h v}} = {{x}\over{y}} 
\EE
with many alternative possibilities. If we require a non-zero 
limit of $\varphi$ this amounts to an asymptotic non-zero value of $x$. If this 
value is finite, say $x=x_o$ we get asymptotically
\BE
\label{finitephi}
    \varphi \to v \tanh(x_o)
\EE
and
\BE
\label{finiteg}
G_J(0) \to  {{ y}\over{M^2_h \cosh^2(x_o)}} \to \infty
\EE
implying the existence of massless modes for every value of $\varphi$. 
On the other hand, for $ x \to \pm \infty$ we obtain
\BE
\varphi \to \pm v
\EE
and in this case $G_J(0)$ tend to 
${{1}\over{M^2_h}}$ or to $+\infty$ depending on whether $y$
diverges slower or faster than $\cosh^2(x)$. 

The above results admit a simple
geometrical interpretation in terms of the shape of the
effective potential $V_{\rm LT}(\varphi)$. 
After obtaining $J$ as a function of
$\varphi$ from Eq.(\ref{phij2}), the inverse zero-momentum propagator
in a given background $\varphi$ is related to the second-derivative of the 
Legendre-transformed effective potential, namely
\BE
\label{step1}
G^{-1}_\varphi(0)= {{dJ}\over{d\varphi}}=
       \frac{ d^2 V_{\rm LT}}{d \varphi^2} 
\EE
In this case, Eqs.(\ref{finitephi}) and (\ref{finiteg}) require
a vanishing result from Eq.(\ref{step1}) when $ -v < \varphi < v$. This 
is precisely what happens since the Legendre-transformed effective potential
becomes flat in the region enclosed by the absolute minima of the non-convex
effective potential when
$\Omega \to \infty$. This is the usual `Maxwell construction' where
$V_{\rm LT}(\varphi)=V_{\rm NC}(\pm v)$, 
for $-v \leq \varphi \leq v$, and 
$ V_{\rm LT}(\varphi)=V_{\rm NC}(\varphi)$ for $\varphi^2 > v^2$.

Notice, however, that the limit of (\ref{step1}) for $\varphi \to \pm v$
is different from (\ref{newident}). In fact, identifying 
the inverse propagators in Eqs. (\ref{newident}) and (\ref{step1}) amounts 
to a much stronger assumption: 
the derivative in Eq.(\ref{step1}) has to be a left- (or right-)
derivative depending on whether we consider the point $\varphi=-v$ 
( or $\varphi=+ v$). However, this is just a prescription since 
derivatives depend on the chosen path unless one deals with infinitely 
differentiable functions. 

Therefore, in general, 
Eq.(\ref{step1}) leads to multiple
solutions for the inverse propagator at the absolute minima $\pm v$. Indeed, 
one can define an exterior derivative for which
$G^{-1}_{\rm ext}(0)=M^2_h$ but one also finds
$G^{-1}_{\rm int}(0)=0$, as when approaching the points
$\pm v$ from the internal region 
where the Legendre-transformed potential becomes flat for $\Omega \to \infty$.
 These two different alternatives
correspond to the various 
limits $y \to \infty$ and $x \to \pm \infty$ in Eq.(\ref{GJ2}) such that
${{y}\over{\cosh^2(x)}}$ tends to zero or infinity.

Now, although determining the finite-momentum propagator requires 
to study the response to a space-time dependent source, our results allow 
to draw some definite conclusions. In the broken phase, i.e.
for all values of $\varphi$ that remain non-vanishing when $J \to \pm 0$, 
there are solutions with ${G}^{-1}_\varphi(0)=0$. 
Therefore, there will always be solutions
${G}^{-1}_\varphi(p)$ that vanish when $p_\mu \to 0$ implying the existence
of gap-less modes whose energy 
also vanishes when ${\bf{p}} \to 0$. We can express 
the required relations for such modes
in terms of unknown slope parameters $\eta$ 
\BE
\label{eta}
    \lim_{ {\bf{p}}\to 0 }
 \tilde{E}^2({\bf{p}})=\eta 
{\bf{p}}^2~~~~~~~~~~~~~~~
\EE
These describe long-wavelength 
collective excitations and are a direct 
consequence of the quantum nature of the `Higgs condensate'.
 
Quite independently of the presence of
massive modes, the massless modes dominate the propagation over large 
distances. In the spontaneously broken phase they are clearly visible
in the lattice data of ref.\cite{cea3} where the energy-gap $\tilde{E}(0)$ 
is found to 
become smaller and smaller by increasing the lattice size. As such, 
$\tilde{E}(0)$ 
cannot represent the operative definition of the Higgs mass $M_h$. This
has to be extracted from the data at larger $|{\bf{p}}|$ that are well reproduced
by the single-particle form
$\sqrt{  {\bf{p}}^2 + {\rm const} }$ after checking that
the numerical value of the constant 
under square root does not depend on the lattice size \cite{cea2,cea3}.
As mentioned at the beginning, insisting in the identification 
$\tilde{E}(0)=M_h$ would represent
the relativistic analogue of applying Landau's definition
of a quasi-particle mass in a system that has a phonon excitation branch for 
${\bf{p}} \to 0$.

We conclude remarking that the peculiar infrared
behaviour we have pointed out does not depend at any stage on the existence
of a continuous symmetry of the classical potential. 
As such, there should be no differences in a spontaneously
broken O(N) theory. Beyond the approximation where
the `Higgs condensate' is treated as a classical
c-number field, one has to 
perform one more integration over the zero-momentum mode of the condensed
$\sigma-$field. Therefore, one finds massless solutions
by computing the inverse propagator of the
$\sigma-$field through Eq.(\ref{step1}) . In this sense, the
existence of gap-less modes of the singlet Higgs field does not depend
on the number of field components.

\end{document}